\documentclass{aa}  

\usepackage{graphicx}
\usepackage{txfonts}
\usepackage{pifont}
\newcommand{\cmark}{\ding{51}}%
\newcommand{\xmark}{\ding{55}}%

\usepackage{subfigure}
\usepackage{multirow}
\usepackage{cancel}

\usepackage{array} 

\RequirePackage[colorlinks,citecolor=blue,urlcolor=blue,linkcolor=blue]{hyperref}

\usepackage[normalem]{ulem}

\begin{document}

   \title{Beyond the Hellings-Downs curve}

   \subtitle{Non-Einsteinian gravitational waves in pulsar timing array correlations}

   \author{Reginald Christian Bernardo
          \inst{1,2}\fnmsep\thanks{\email{reginald.bernardo@apctp.org}}
          \and
          Kin-Wang Ng\inst{2,3}\fnmsep\thanks{\email{nkw@phys.sinica.edu.tw}}
          }

   \institute{Asia Pacific Center for Theoretical Physics, Pohang 37673, Republic of Korea 
         \and
             Institute of Physics, Academia Sinica, Taipei 11529, Taiwan
         \and
             Institute of Astronomy and Astrophysics, Academia Sinica, Taipei 11529, Taiwan \\
             }


 
  \abstract
    {The recent astronomical milestone by the pulsar timing arrays (PTA) has revealed galactic-size gravitational waves (GW) in the form of a stochastic gravitational wave background (SGWB), correlating the radio pulses emitted by millisecond pulsars. This draws the outstanding questions toward the origin and the nature of the SGWB; the latter is synonymous to testing how quadrupolar the inter-pulsar spatial correlation is. In this paper, we tackle the nature of the SGWB by considering correlations beyond the Hellings-Downs (HD) curve of Einstein's general relativity. We put the HD and non-Einsteinian GW correlations under scrutiny with the NANOGrav and the CPTA data, and find that both data sets allow a graviton mass $m_{\rm g} \lesssim 1.04 \times 10^{-22} \ {\rm eV}/c^2$ and subluminal traveling waves. We discuss gravitational physics scenarios beyond general relativity that could host non-Einsteinian GW correlations in the SGWB and highlight the importance of the cosmic variance inherited from the stochasticity in interpreting PTA observation.}


   \maketitle
%

\begin{figure}[t]
    \centering
    \includegraphics[width = 0.4\textwidth]{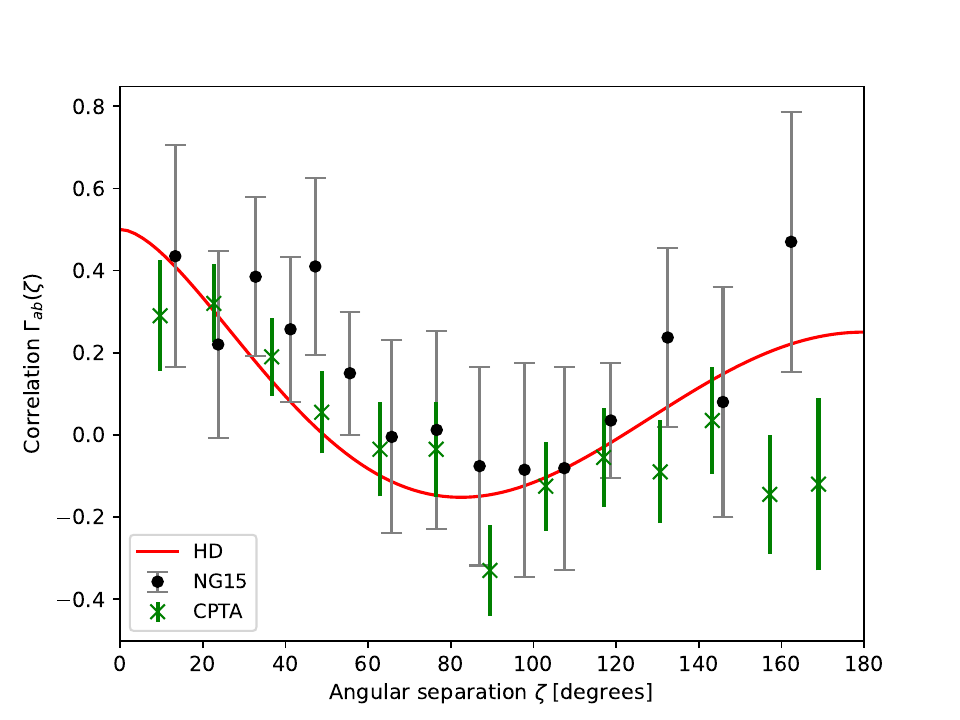}
    \caption{Hellings-Downs curve. The HD curve (red line) with the NANOGrav (\cite{NANOGrav:2023gor}) (black dots) and CPTA (\cite{Xu:2023wog}) (green crosses) inter-pulsar correlation samples shows compelling support for the stochastic gravitational wave background in the galaxy.}
    \label{fig:HD}
\end{figure}

The marvelous detection of the stochastic gravitational wave background (SGWB) by the pulsar timing array (PTA) astronomical community (\cite{NANOGrav:2023gor, Reardon:2023gzh, EPTA:2023fyk, Xu:2023wog}) is distinct in various ways compared with the previous observations of gravitational waves (\cite{KAGRA:2021vkt, Romano:2023zhb}). Conceptually, since the SGWB is light-years long in size, the PTAs have had to monitor millisecond pulsars over the span of years to decades (\cite{Detweiler:1979wn, Romano:2016dpx, Burke-Spolaor:2018bvk, NANOGrav:2020spf}), and overcome challenges related to sustaining such an illustrious effort. The astronomical milestone most importantly shows gravitational waves (GW) manifest the very feature that sets apart waves---`interference'---to produce a galactic superposition that spatially correlates the time-of-arrival of radio pulses of millisecond pulsars. This correlation, embodied by the Hellings-Downs curve (\cite{Hellings:1983fr, Jenet:2014bea}) (Fig. \ref{fig:HD}), underlines a pronounced quadrupolar shape that distinguishes the presence of a gravitational wave background.

The resolution of this elusive signal is the culmination of decades of theoretical and experimental work in PTA science (\cite{Phinney:2001di, 2010CQGra..27h4013H, Lee:2010cg, Lee:2011et, Dai:2012bc, Yunes:2013dva, Gair:2014rwa, Vigeland:2013zwa, Lommen:2015gbz, Vagnozzi:2020gtf, Taylor:2021yjx, Pol:2022sjn, Allen:2022dzg, Bernardo:2022xzl, Bernardo:2023bqx, Allen:2022ksj, Allen:2023kib}). Now, the nanohertz GW window brings forth GW astronomy to its multiband era (\cite{Lasky:2015lej, 2021NatRP...3..344B, Lambiase:2023pxd}), giving PTAs their invaluable edge for GW and multimessenger science. But even in the nanohertz GW regime alone, the questions in PTA science has, since the observation of the HD curve, transformed to the source and the nature of the SGWB, keeping physicists and astronomers at the edge of their seats.

\begin{table*}[ht]
\centering
\scriptsize
\renewcommand{\arraystretch}{1.2} 
\caption{Constraints on non-Einsteinian GW polarizations using NANOGrav (\cite{NANOGrav:2023gor}) and Chinese PTA (\cite{Xu:2023wog}) data.}
\begin{tabular}{|c r|c r|r|r|r|r|r|}
\hline
GW Mode/s & Data \ & CV & Speed $v/c$ \ & \ Mass \ & \ $r^2$ \ & \ $\overline{\chi}^{2}$ \ & \ $\log {\cal B}_{\rm GRN}$ \ & \ $\log {\cal B}_{\rm HD}$ \ \\ \hline 
\multirow{3}{*}{HD}
& NG15 \ & & $1$ \ & $0$ \ & $0$ \ & \ $0.64$ \ & \ {$5.8 \pm 0.1$} \ & {$0$} \ \\
& CPTA \ & & $1$ \ & $0$ \ & $0$ \ & \ $1.29$ \ & \ {$5.6 \pm 0.2$} \ & {$0$} \ \\
& \ NG15+CPTA \ & & $1$ \ & $0$ \ & $0$ \ & \ $0.94$ \ & \ {$9.9 \pm 0.3$} \ & {$0$} \ \\
\hline
\multirow{6}{*}{Tensor}
& NG15 \ & & \ $> 0.38$ \ & \ $\lesssim 1.21$ \ & \ 0 \ & \ $0.54$ \  & \ $5.7 \pm 0.1$ \ & {$-0.10 \pm 0.01$} \ \\
& NG15 \ & \cmark \ \ \ & \ $> 0.42$ \  & \ $\lesssim 1.19 $ \ & \ 0 \ & \ $0.56$ \ & \ $6.0\pm 0.1$ \ & {$0.23 \pm 0.01$} \ \\
& CPTA \ & & \ $> 0.46$ \ & \ $\lesssim 1.16 $ \ & \ 0 \ & \ $1.03$ \ & \ $6.1\pm 0.2$ \ & {$0.44 \pm 0.01$} \ \\
& CPTA \ & \cmark \ \ \ & \ $> 0.52$ \  & \ $\lesssim 1.12 $ \ & \ 0 \ & \ $0.70$ \ & \ $8.8\pm 0.2$ \ & {$3.16 \pm 0.01$} \ \\
& \ NG15+CPTA \ & & \ $> 0.57$ \ & \ $\lesssim 1.08 $ \ & \ 0 \ & \ $0.86$ \ & \ $10.3 \pm 0.3$ \ & {$0.41 \pm 0.01$} \ \\
& \ NG15+CPTA \ & \cmark \ \ \ & \ $> 0.61$ \ & \ $\lesssim 1.04 $ \ & \ 0 \ & \ $0.62$ \ & \ $13.4\pm 0.3$ \ & {$3.48 \pm 0.02$} \ \\
\hline 
\multirow{6}{*}{Vector}
& NG15 \ & & \ $0.53^{+0.29}_{-0.13}$ \ & \ $\simeq (0.75, 1.20) $ \ & \ 0 \ & \ $0.50$ \  & \ $6.0 \pm 0.1$ \ & {$0.20 \pm 0.02$} \ \\
& NG15 \ & \cmark \ \ \ & \ $0.58^{+0.29}_{-0.12}$ \  & \ $\simeq (0.66, 1.17) $ \ & \ 0 \ & \ $0.44$ \ & \ $6.5 \pm 0.1$ \ & {$0.71 \pm 0.02$} \ \\
& CPTA \ & & \ $0.49^{+0.23}_{-0.10}$ \ & \ $\simeq (0.92, 1.21) $ \ & \ 0 \ & \ $1.04$ \ & \ $6.3 \pm 0.2$ \ & {$0.65 \pm 0.02$} \ \\
& CPTA \ & \cmark \ \ \ & \ $0.60^{+0.25}_{-0.10}$ \  & \ $\simeq (0.70, 1.14) $ \ & \ 0 \ & \ $0.56$ \ & \ $9.4 \pm 0.2$ \ & {$3.78 \pm 0.02$} \ \\
& \ NG15+CPTA \ & & \ $0.55^{+0.17}_{-0.06}$ \ & \ $\simeq (0.91, 1.14) $ \ & \ 0 \ & \ $0.76$ \ & \ $11.1 \pm 0.3$ \ & {$1.15 \pm 0.03$} \ \\
& \ NG15+CPTA \ & \cmark \ \ \ & \ $0.66^{+0.16}_{-0.07}$ \ & \ $\simeq (0.74, 1.05)$ \ & \ 0 \ & \ $0.50$ \ & \ $14.8 \pm 0.3$ \ & {$4.86 \pm 0.03$} \ \\
\hline 
\multirow{6}{*}{HD-V}
& NG15 \ & & \ $< 0.53$ \ & \ --- \ & \ $< 0.56$ \ & \ $0.54$ \  & \ $5.6 \pm 0.1$ \ & {$-0.16\pm 0.01$} \ \\
& NG15 \ & \cmark \ \ \ & \ $0.43^{+0.18}_{-0.39}$ \  & \ --- \ & \ $<0.61$ \ & \ $0.54$ \ & \ $6.1 \pm 0.1$ \ & {$0.34 \pm 0.01$} \ \\
& CPTA \ & & \ $< 0.54$ \ & \ --- \ & \ $< 0.14$ \ & \ $1.29$ \ & \ $3.5 \pm 0.2$ \ & {$-2.12 \pm 0.03$} \ \\
& CPTA \ & \cmark \ \ \ & \ $< 0.57$ \  & \ --- \ & \ $< 0.29$ \ & \ $0.72$ \ & \ $7.9 \pm 0.2$ \ & {$2.32 \pm 0.02$} \ \\
& \ NG15+CPTA \ & & \ $< 0.52$ \ & \ --- \ & \ $< 0.14$ \ & \ $0.94$ \ & \ $7.9 \pm 0.3$ \ & {$-1.99 \pm 0.03$} \ \\
& \ NG15+CPTA \ & \cmark \ \ \ & \ $< 0.56$ \ & \ --- \ & \ $< 0.28$ \ & \ $0.64$ \ & \ $12.8 \pm 0.3$ \ & {$2.86\pm 0.02$} \ \\
\hline 
\multirow{6}{*}{HD-$\phi$}
& NG15 \ & & \ \xmark \ \ & \ --- \ & \ $0.33 \pm 0.14$ \ & \ $0.26$ \  & \ $7.5 \pm 0.1$ \ & {$1.68 \pm 0.02$} \ \\
& NG15 \ & \cmark \ \ \ & \ \xmark \ \ & \ --- \ & \ $0.55^{+0.23}_{-0.29}$ \ & \ $0.14$ \ & \ $9.1 \pm 0.1$ \ & {$3.29 \pm 0.01$} \ \\
& CPTA \ & & \ \xmark \ \ & \ --- \ & \ $< 0.05$ \ & \ $1.29$ \ & \ $2.7 \pm 0.2$ \ & {$-2.96 \pm 0.05$} \ \\
& CPTA \ & \cmark \ \ \ & \ \xmark \ \  & \ --- \ & \ \xmark \ \ & \ $0.68$ \ & \ $8.9 \pm 0.2$ \ & {$3.27 \pm 0.01$} \ \\
& \ NG15+CPTA \ & & \ \xmark \ \ & \ --- \ & \ $< 0.08$ \ & \ $0.93$ \ & \ $7.6 \pm 0.3$ \ & {$-2.32 \pm 0.04$} \ \\
& \ NG15+CPTA \ & \cmark \ \ \ & \ \xmark \ \ & \ --- \ & \ $> 0.49$ \ & \ $0.44$ \ & \ $16.4 \pm 0.3$ \ & {$6.51 \pm 0.01$} \ \\
\hline
\multirow{6}{*}{T-$\phi$} \
& NG15 \ & & \ \xmark \ \ & \ \xmark \ \ & \ $0.36 \pm 0.14$ \ & \ $0.21$ \  & \ $7.7\pm 0.1$ \ & {$1.94 \pm 0.02$} \ \\
& NG15 \ & \cmark \ \ \ & \ \xmark \ \  & \ \xmark \ \ & \ $0.55^{+0.23}_{-0.28}$ \ & \ $0.11$ \ & \ $9.2 \pm 0.1$ \ & {$3.40 \pm 0.01$} \ \\
& CPTA \ & & \ $> 0.46$ \ \ & \ $\lesssim 1.16$ \ & \ $< 0.07$ \ & \ $1.14$ \ & \ $3.5 \pm 0.2$ \ & {$-2.14 \pm 0.04$} \ \\
& CPTA \ & \cmark \ \ \ & \ \xmark \ \  & \ \xmark \ \ & \ $> 0.54$ \ & \ $0.58$ \ & \ $9.1 \pm 0.2$ \ & {$3.49 \pm 0.01$} \ \\
& \ NG15+CPTA \ & & \ $> 0.56$ \ \ & \ $\lesssim 1.08$ \ & \ $< 0.11$ \ & \ $0.83$ \ & \ $8.8 \pm 0.3$ \ & {$-1.14 \pm 0.04$} \ \\
& \ NG15+CPTA \ & \cmark \ \ \ & \ \xmark \ \ & \ \xmark \ \ & \ $> 0.53$ \ & \ $0.40$ \ & \ $16.9\pm 0.3$ \ & {$6.93 \pm 0.01$} \ \\\hline
\end{tabular}
\tablefoot{HD: Hellings-Downs curve; `Tensor'/`Vector': subluminal GWs; HD-$\phi$/V: luminal tensor GWs + scalar/vector GWs with speed $v$ and fraction $r^2$; T-$\phi$: tensor and scalar GWs with the same speed. `\cmark' in CV cells: cosmic variance included (\cite{Allen:2022dzg, Bernardo:2022xzl}). `\xmark'/`---': unconstrained/not applicable parameters. $\overline{\chi}^2$: reduced chi-squared; {$\log {\cal B}_{\cal H}$: Bayes factor vs. null hypothesis ${\cal H}$ (Gaussian random noise/GRN or HD)
; $\log {\cal B}_{\cal H} > (<) \ 0$: evidence for (against) correlations model. Masses in units of $10^{-22}$ eV/$c^2$.} Data: NANOGrav (NG15), Chinese PTA (CPTA).}
\label{tab:summary}
\end{table*}

A most natural source of the SGWB are astrophysical supermassive black hole binaries, or rather the nanohertz GWs they emit (\cite{Rajagopal:1994zj, Jaffe:2002rt, Sesana:2008mz, Ravi:2014nua, Shannon:2015ect, Mingarelli:2017fbe, Liu:2021ytq, NANOGrav:2021ini, NANOGrav:2023pdq, EPTA:2023gyr, Ellis:2023dgf, Cannizzaro:2023mgc}); however, the present data allow for a more speculative and interesting possibilities of cosmological origin such as primodrial black holes, first order phase transitions, and domain walls (\cite{Chen:2019xse, NANOGrav:2021flc, Xue:2021gyq, Moore:2021ibq, NANOGrav:2023hvm, EPTA:2023xxk, Vagnozzi:2023lwo, Franciolini:2023wjm, Bai:2023cqj, Megias:2023kiy, Jiang:2023gfe, Zhang:2023nrs, Figueroa:2023zhu, Bian:2023dnv, Niu:2023bsr, Depta:2023qst, Abe:2023yrw, Servant:2023mwt, Ellis:2023oxs, Bhaumik:2023wmw, Ahmed:2023pjl, Antusch:2023zjk, Aghaie:2023lan, Konoplya:2023fmh, Basilakos:2023xof, Basilakos:2023jvp, Ben-Dayan:2023lwd, Ahmadvand:2023lpp, Choudhury:2023hfm}). These different early universe high energy physics sources are distinguished by the frequency spectrum. On the other hand, the gravitational nature of the SGWB is perceptible through the inter-pulsar spatial correlations (\cite{Chamberlin:2011ev, Qin:2018yhy, Qin:2020hfy, Ng:2021waj, Chen:2021wdo, Liu:2022skj, Hu:2022ujx, Bernardo:2022rif}). The quadrupolar shape of the HD curve is inherited from the tensorial nature of the GW polarizations that are now so familiar from ground-based GW observations. In this context, the HD curve can be viewed as a reference point for testing gravity in the nanohertz GW regime where the prospects are as rich as the sources (\cite{Liang:2021bct, Bernardo:2022vlj, Bernardo:2023mxc, Wu:2023pbt, Liang:2023ary, Bernardo:2023pwt}).

In line with this gravitational physics theme, we perform an exhaustive phenomenological search for non-Einsteinian GWs (Tab. \ref{tab:summary}) using the NANOGrav (\cite{NANOGrav:2023gor}) and the CPTA (\cite{Xu:2023wog}) inter-pulsar correlations data, with the following physical premise: that the HD correlation is a byproduct of general relativity (GR); then, the alternative hypothesis is that observed departures from the HD curve hint a nanohertz deviation from GR. In doing so, we take into account the mean of the expected correlation signal of the SGWB and its cosmic variance (CV). The theoretical uncertainty owed due to the stochasticity of the signal is a relatively recent consideration, first brought up a year ago for the HD correlation (\cite{Allen:2022dzg}) and was later extended to subluminal and non-Einsteinian GWs (\cite{Bernardo:2022xzl}). As with cosmic microwave background (CMB) measurements, the CV of PTA correlation measurements of the SGWB provides a nature inherent leeway for alternative models at large scales, regardless of the precision of the experiment, and plays a key role in testing gravity.


Nanohertz GR deviations can hail from subluminal GW propagation and/or non-GR gravitational degrees of freedom (d.o.f.). Subluminal GW propagation can come from a massive graviton or from modifying the dispersion relation of GWs (\cite{Baskaran:2008za, Lee:2013awh}); the latter may be more natural from an effective field theory point of view. This manifests an enhanced quadrupolar power compared with its luminal counterpart that give the HD correlation (\cite{Mihaylov:2019lft, Qin:2020hfy, Bernardo:2022rif}). In the same vein, non-Einsteinian GW polarizations are brought by scalar and vector gravitational d.o.f.s that are tied to alternative theories of gravity (\cite{Clifton:2011jh, Joyce:2014kja, Nojiri:2017ncd, Cornish:2017oic, Kase:2018aps, Ferreira:2019xrr, OBeirne:2019lwp, Battista:2021rlh, Battista:2022hmv, Odintsov:2023weg, Tasinato:2023zcg}). The wiggle room around GW and local constraints of such GW polarizations is that nanohertz GWs propagate in dynamical regions in the galaxy where screening mechanisms may not necessarily take place. More concretely, screening effects often go with a quasistatic consideration (\cite{Hui:2009kc, Wang:2012kj, deRham:2012az, Brax:2011sv, Ali:2012cv, Andrews:2013qva}) that galactic long oscillatory GWs do not have to satisfy. Non-Einsteinian GW modes can be constrained through their nontrivial monopolar and dipolar correlation signals with specific values controlled by the their propagation. The modelling of the non-HD correlation signals are disclosed in more detail in the {\bf Methods} Appendix, relying on so-called harmonic analysis or power spectrum approach.

\begin{figure}[t]
    \centering    \includegraphics[width = 0.4\textwidth]{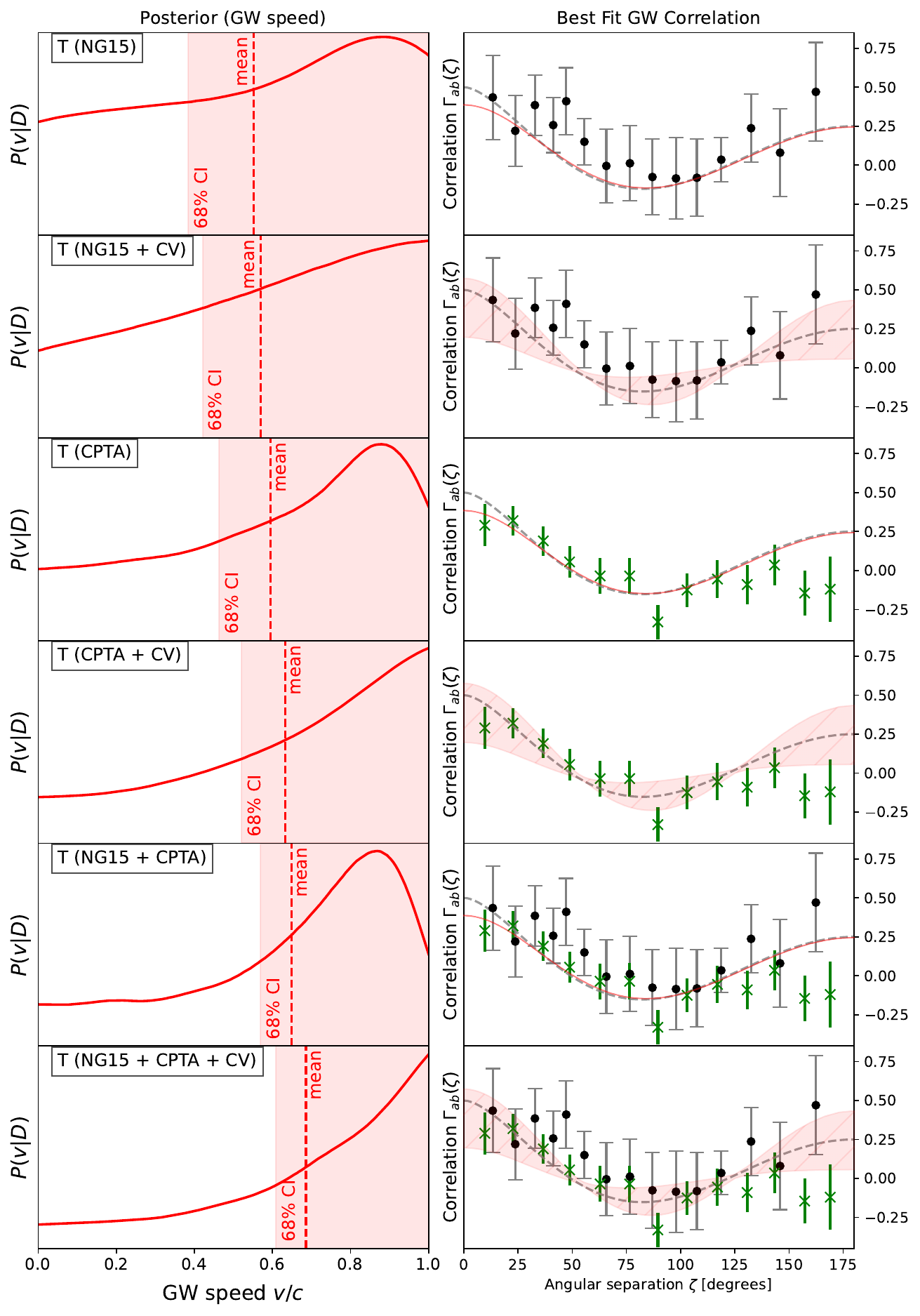}
    \caption{Subluminal GWs. [left] Constraints on the GW speed $v$ for tensor modes using the NANOGrav (\cite{NANOGrav:2023gor}) (black dots) and CPTA (\cite{Xu:2023wog}) (green crosses) correlations data. [right] Best fit correlation samples (red curve/$1\sigma$-bands) corresponding to the left panel. The HD curve (gray dashed) is shown for reference.}
    \label{fig:tensorconstraints}
\end{figure}

\begin{figure}[t]
    \centering    \includegraphics[width = 0.4\textwidth]{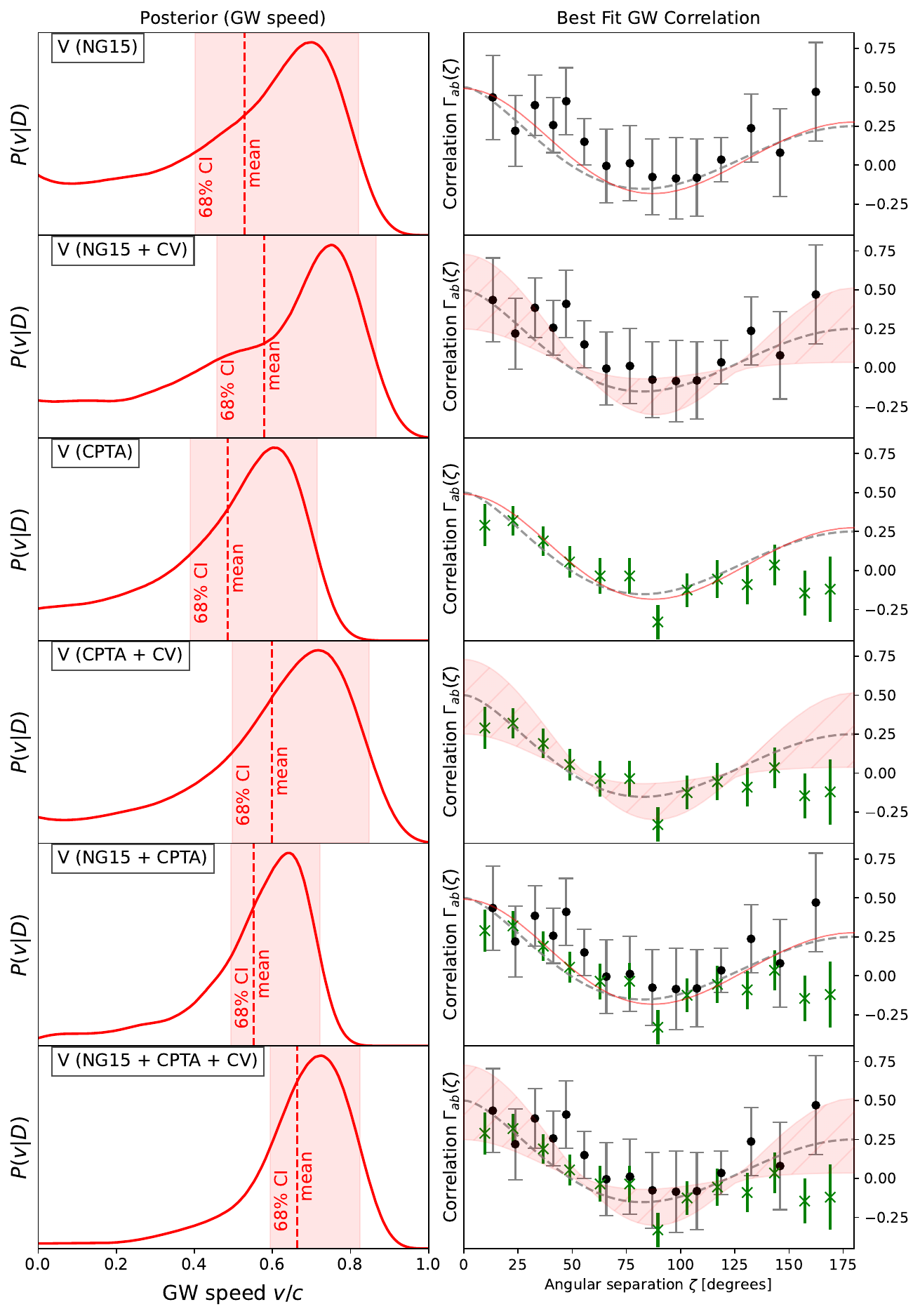}
    \caption{Subluminal GWs. [left] Constraints on the GW speed $v$ for vector modes using the NANOGrav \cite{NANOGrav:2023gor} (black dots) and CPTA \cite{Xu:2023wog} (crosses) correlations data. [right] Best fit correlation samples (red curve/$1\sigma$-bands) corresponding to the left panel. The HD curve (gray dashed) is shown for reference.}
    \label{fig:vectorconstraints}
\end{figure}

\begin{figure}[t]
    \centering    \includegraphics[width = 0.4\textwidth]{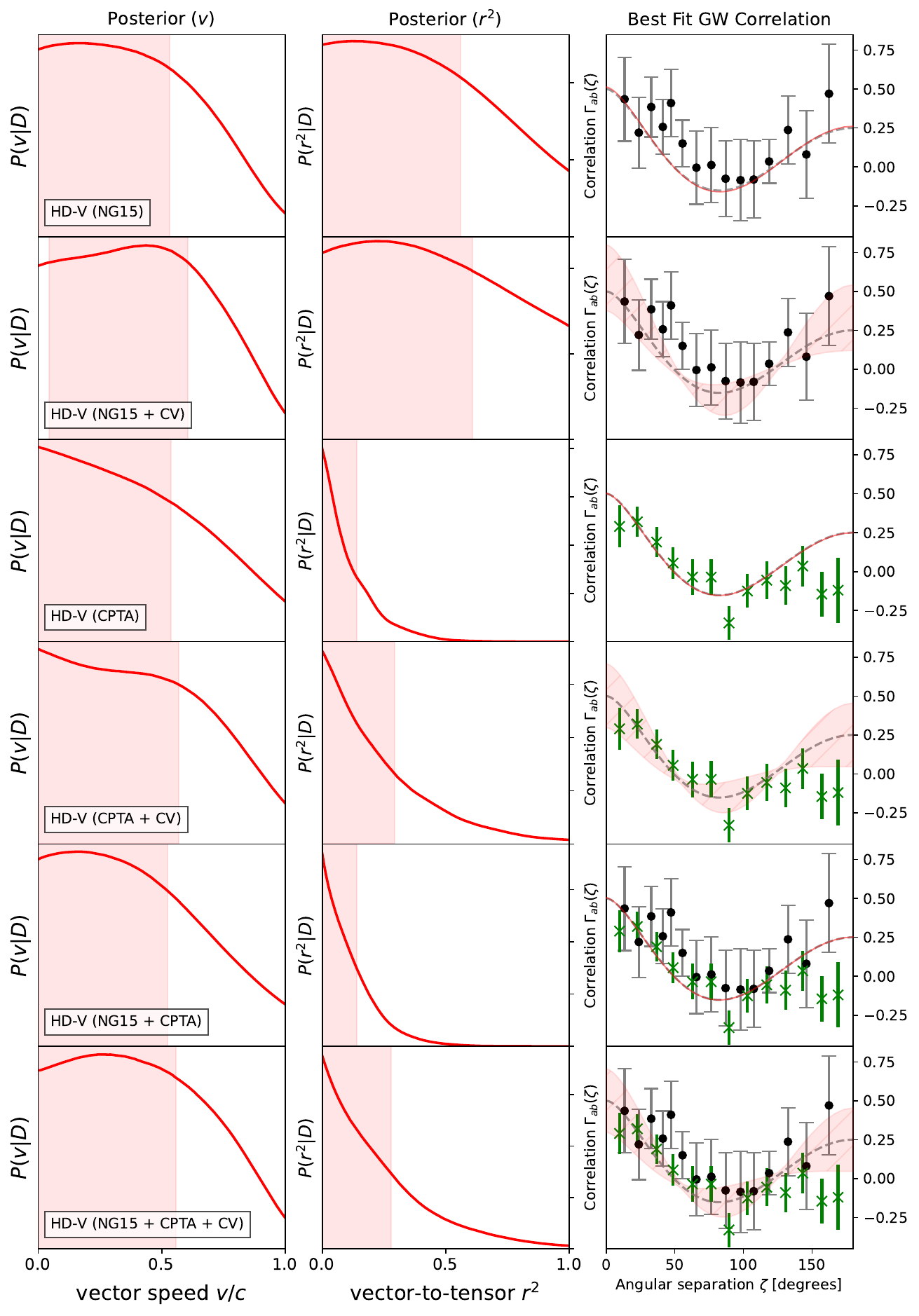}
    \caption{HD-V. Constraints on the [left] speed $v$ and [middle] vector-to-tensor fraction $r^2$ for vector modes mixing with luminal GWs (HD) in the SGWB using the NANOGrav (\cite{NANOGrav:2023gor}) (black dots) and CPTA (\cite{Xu:2023wog}) (green crosses) correlations data. [right] Best fit correlation samples (red curve/$1\sigma$-bands) corresponding to the left and middle panels. The HD curve (gray dashed) is shown for reference.}
    \label{fig:HDVconstraints}
\end{figure}

\begin{figure}[t]
    \centering    \includegraphics[width = 0.4\textwidth]{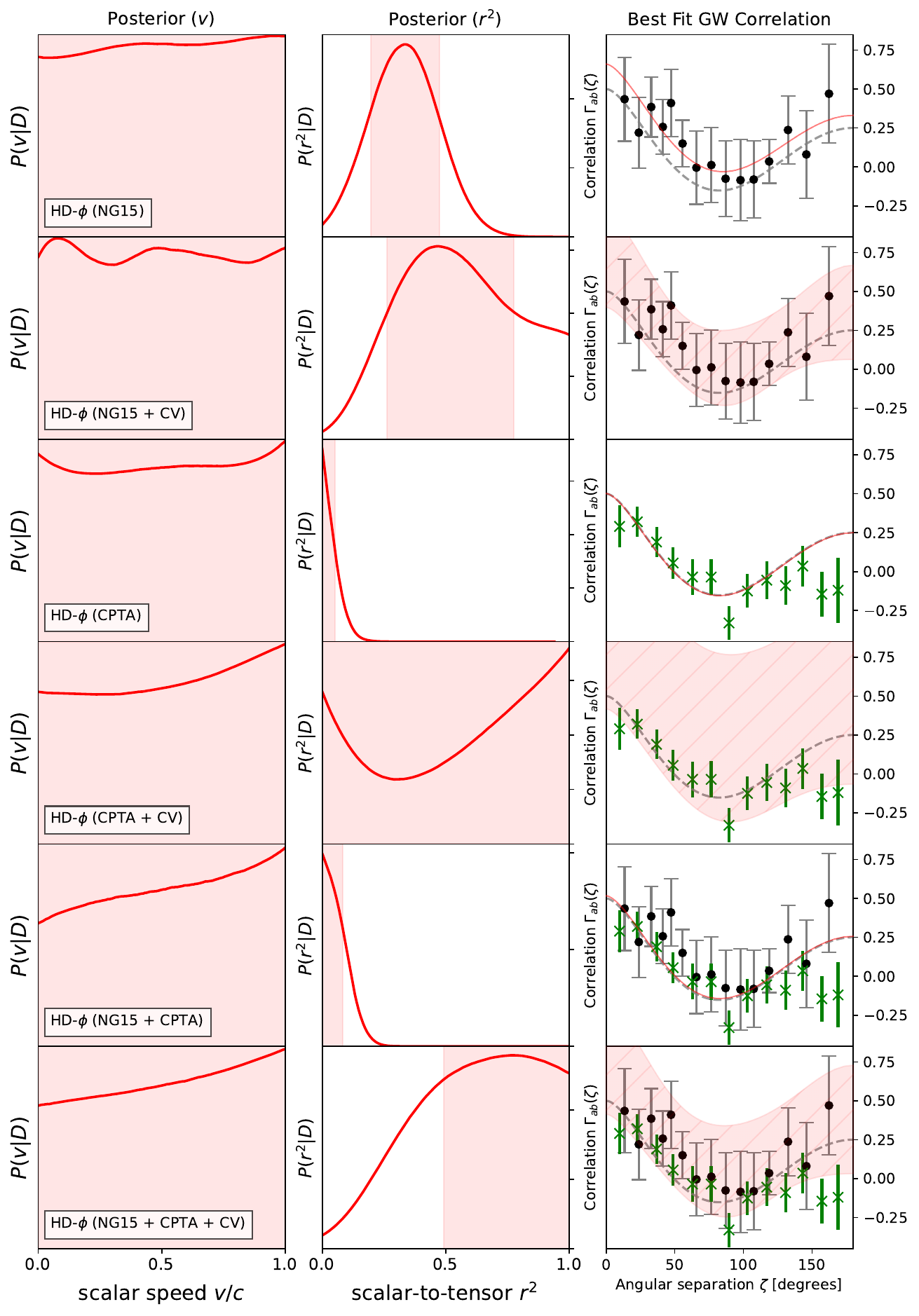}
    \caption{HD-$\phi$. Constraints on the [left] speed $v$ and [middle] scalar-to-tensor fraction $r^2$ for scalar modes mixing with luminal GWs (HD) in the SGWB using the NANOGrav (\cite{NANOGrav:2023gor}) (black dots) and CPTA (\cite{Xu:2023wog}) (green crosses) correlations data. [right] Best fit correlation samples (red curve/$1\sigma$-bands) corresponding to the left and middle panels. The HD curve (gray dashed) is shown for reference.}
    \label{fig:HDphiconstraints}
\end{figure}

\begin{figure}[t]
    \centering    \includegraphics[width = 0.4\textwidth]{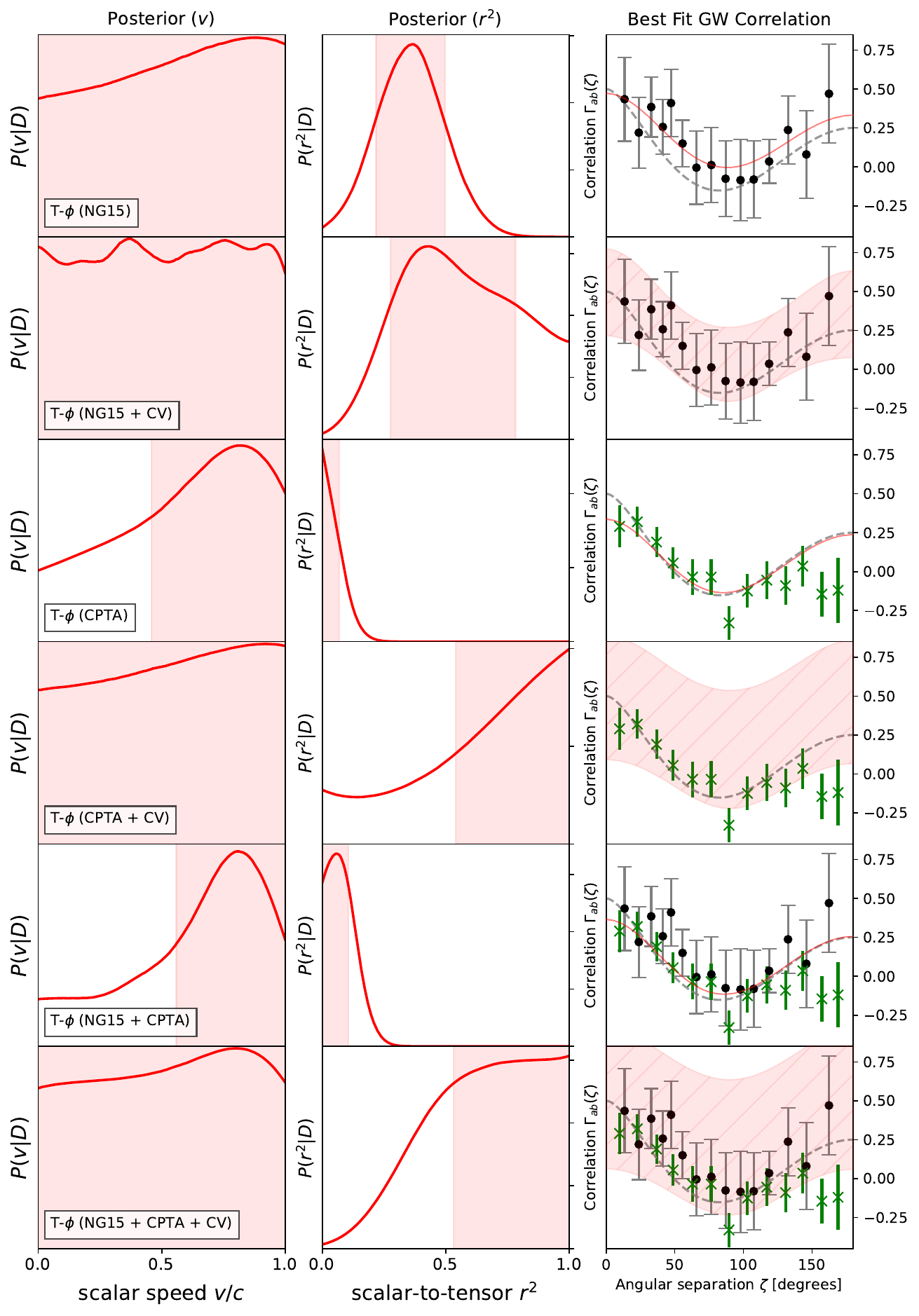}
    \caption{T-$\phi$. Constraints on the [left] GW speed $v$ and [middle] scalar-to-tensor fraction $r^2$ for scalar and tensor modes in the SGWB using the NANOGrav (\cite{NANOGrav:2023gor}) (black dots) and CPTA (\cite{Xu:2023wog}) (green crosses) correlations data. [right] Best fit correlation samples (red curve/$1\sigma$-bands) corresponding to the left and middle panels. The HD curve (gray dashed) is shown for reference.}
    \label{fig:Tphiconstraints}
\end{figure}

Our results is summarized in Tab. \ref{tab:summary} and discussed for the rest of this paper. Each row in Tab. \ref{tab:summary} corresponds to a row in Figures \ref{fig:tensorconstraints}-\ref{fig:Tphiconstraints} that illustrates the constrained parameter space for a fixed GW polarization and the best fit to the correlation data points. For subluminal tensor (T) and vector (V) GWs, the parameter is the GW speed, $v$, interpreted with a graviton mass. For the tensor-non-tensor GWs, an extra parameter $r^2$ comes to play where $r$ is a phenomenological parameter that measures the amplitude ratio of the non-tensorial modes to the tensorial ones. In the figures, the error bars/bands stand for the 68\% confidence limits, and the black dots and green crosses correspond to the NANOGrav and the CPTA data, respectively. The best fit GW correlation curves are given as a red line or a red-hatched band representing the $1\sigma$-uncertainty due to the CV whenever the theoretical variance is included in the analysis. The HD curve (gray dashed) is presented in all the correlation plots for reference. In the presentation, we align with standard procedure for detecting spatial correlation in PTA data by weighting evidences against a spatially uncorrelated one. {In addition, we present the Bayes factors of models with non-Einsteinian polarizations versus the one without, aka. the HD curve.}


We take the discussion from the top of Tab. \ref{tab:summary} which is that of subluminal traveling tensor GWs. In this case, the gravitational distinction from the HD correlation is that the tensor GW polarizations move slower in vacuum compared to light. The constraints are presented in Fig. \ref{fig:tensorconstraints}. These boil down to subluminal tensor GWs, $v/c \gtrsim 0.4$, to be a viable alternative to the HD ($v/c = 1$) as the predominant contribution to the SGWB. The result means that the present correlation measurements allow a perfect quadrupole, $\Gamma_{ab}(\zeta) \propto \cos (2\zeta)$, to fit reasonably within the error bars. The physics behind-the-scenes is that the GW correlation becomes more quadrupolar compared with the HD curve the slower GWs become relative to light (\cite{Qin:2020hfy, Bernardo:2022rif}). It should be emphasized that the HD curve ($v/c = 1$) remains a perfectly viable model of the observed correlations. The results enrich this point of view by seeing the HD curve as a particular case of GWs that is tensorial and luminal.

Figure \ref{fig:tensorconstraints} highlights that the constraints on the GW speed gradually tightens when the CV is considered in the parameter estimation (\cite{Allen:2022dzg, Bernardo:2022xzl, Bernardo:2023bqx}). In addition, the results give weight to a joint analysis of the correlations to constrain gravity, in the sense that the constraints tighten up when the data from both PTAs are utilized. In this way, the tightest constraint $v/c > 0.61$ on the speed of nanohertz GWs can be derived when both the NANOGrav and CPTA correlation samples are taken into account. A lower bound to the GW speed translates to an upper bound to the graviton mass: $m_{\rm g} \lesssim 1.04 \times 10^{-22} \ {\rm eV}/c^2$. This keeps the HD curve ($v/c = 1$ or $m_{\rm g} = 0$) a compelling model of the observed SGWB but so are correlations due to subluminal tensor GWs ($v/c < 1$ or $m_{\rm g} \neq 0$). It is worth noting that all the quadrupolar GW correlations are more prominent in the data compared with an uncorrelated process (rightmost column of Tab. \ref{tab:summary} tell the evidence). This is a testament to the presence of the SGWB despite the attendant questions about its source and gravitational nature.

It should be pointed out that subluminal GW propagation in the nanohertz regime remains a perfectly viable astronomical and physical scenario even with a stringent case for luminal GW propagation in the subkiloherz GW band (\cite{deRham:2018red, Bernardo:2023mxc}). Hints to subluminal GW propagation can be found in the measured angular power spectrum of the SGWB (\cite{NANOGrav:2023gor}). For luminal nanohertz GWs that form the HD curve, the multipolar profiles are fashioned in a specific way such that the quadrupole is followed by the octupole and so on, in certain amounts (\cite{Gair:2014rwa, Ng:2021waj}). However, for subluminal GW propagation, the octupole and the higher moments become more suppressed, leading to a dominantly quadrupole corelation profile (\cite{Qin:2020hfy, Bernardo:2022rif}). Direct measurements of the first few power spectrum coefficients can help to distinguish between the HD and subluminal GWs in future data (\cite{Nay:2023pwu}). An endgame to look forward to in PTA science are CV-precise measurements of the correlation power spectrum or the two-point function (\cite{Bernardo:2023pwt}), following in the footsteps of CMB measurements.

Tensor GWs are not unique in its ability to produce a quadrupolar spatial correlation. Vector GWs feature a dominant quadrupole and higher moments, but is mostly characterized by its nonvanishing dipolar power, distinguishing it with tensor GWs (\cite{Qin:2020hfy, Bernardo:2022rif}). The discussion of vector GWs, however, should be taken from an exclusively phenomenological standpoint, for one, because there are not much known astrophysical or cosmological mechanisms that would produce lone vector GWs (\cite{Wu:2023pbt, Bernardo:2023mxc}). Vector modes are furthermore easily diluted by the cosmic expansion. With that in mind, we discuss the phenomenological constraints obtained on vector GWs and the observed correlation of timing residuals of millisecond pulsars (Tab. \ref{tab:summary} `Vector' subrows and Fig. \ref{fig:vectorconstraints}). 

Results tellingly show that vector GWs are a good phenomenological model for the observed spatial correlation for the following reasons. Unlike the tensor modes, it turns out that the data is able to place upper and lower bounds to the GW speed; the upper bound owing to the known divergence of the vector GW correlations in the luminal and infinite distance limit (\cite{NANOGrav:2021ini, Qin:2020hfy, Bernardo:2022rif}). This translates to constraints on the graviton mass should vector GWs be interpreted within the framework of massive gravity (\cite{Bernardo:2023mxc, Wang:2023div}). The vector GW constraints echo two results that were also realized with tensor GWs: the impact of the CV and a joint analysis in narrowing down the parameter space. A subtle point is that vector GW correlations happen to be slightly preferred compared with their tensor counterparts. This apparent contradiction with an earlier result using the NANOGrav 12.5 years data (\cite{Bernardo:2023mxc}) can be attributed to the fact that the present measurements now compellingly show a quadrupolar correlation pattern that was previously hidden away by larger uncertainties.

Once again, since there are no realizable physical processes to let lone vector GWs manifest in the SGWB, however good the fit is, we can not take it beyond a phenomenological interpretation of the measured quadrupolar spatial correlation. Vector GWs nonetheless are a promising avenue for PTA cosmology to showcase its scientific value by closing the window for vector GWs in future iterations of the data. 

The weak field in regions of the galaxy where nanohertz GWs are able to oscillate and superpose is a crucial piece to allow the formation of the SGWB. This same information lets the waves interfere to exhibit the expected light and dark fringes that leave observable traces in the radio signals of the galactic millisecond pulsars. The HD curve and subluminal GWs from lone tensor or vector modes become a physical possibility because of this. Likewise, the same logic anchors the possibility that the SGWB is a mixture of various gravitational components, a prospect that makes both theory and phenomenology of the inter-pulsar correlation more exciting. Our next constraints dwell in this direction: considering SGWB from a dominant tensor GW contribution with subdominant vector/scalar GW components.

A fair starting point is recognizing that the SGWB is quadrupolar, hence tensorial, and that the HD curve is a remarkable fit to the measured spatial correlation. We thus consider GW correlations produced by luminal tensor modes {and} subluminal vector (HD-V) or scalar (HD-$\phi$) modes in the SGWB. The correlations model HD-$\phi$ could be realized in scalar-tensor theory with Gauss-Bonnet couplings constrained in the subkilohertz GW band (\cite{Kim:1999dq, Kim:2000ym, Capuano:2023yyh}). We suspect an analogous vector-tensor theory (\cite{Aoki:2023jvt}) may be able to realize the HD-V model, i.e., luminal tensor and subluminal vector modes. In these mixed nature SGWB scenarios, $v$ describes the speed of the non-luminal modes and $r^2$ is a phenomenological parameter that characterizes the fraction of non-tensorial modes in the SGWB. More concretely, since the mean and the variance of the correlation behave as two- and four-point functions of the GW amplitudes, $\Gamma_{ab} \sim \langle h_a h_b \rangle$ and $\sigma^{\rm CV}_{ab} \sim \langle h_a h_a h_b h_b \rangle$, respectively, it is possible to associate $r$ to the amplitude of non-tensorial modes relative to the tensorial ones (see {\bf Methods} Appendix for details). Following the same context, we consider subluminal tensor and scalar modes with the {same} speed. Our resulting T-$\phi$ correlations model could be tied with massive gravity where the modal speed is controlled by the graviton mass, presuming that the vector modes' contribution is negligible.

The results are shown in the `HD-V', `HD-$\phi$', and `T-$\phi$' rows of Tab. \ref{tab:summary} and Figs. \ref{fig:HDVconstraints}-\ref{fig:Tphiconstraints}, and discussed in what follows.

On additional vector GWs (HD-V), the correlation data sets happen to be able to place upper bounds to the vector modes' speed and fraction in the SGWB. In these cases, the best fits visually reproduce the HD correlation's mean and CV, when applied, as shown in Fig. \ref{fig:HDVconstraints}. This is realized in the constraints by the 68\% confidence upper bounds to the vector GWs fraction parameter obtained in all of the cases. It is also interesting that when the CV is taken into account the parameter constraints become slightly wider, in contrast when only lone tensor or vector GWs are considered. Either way, the NANOGrav and CPTA joint analysis including the CV places an upper bound to the vector modes' speed of $v/c \lesssim 0.56$ and to their fractional contribution to the SGWB of $r^2 < 0.28$. The evidence of the HD-V correlation is a little lower compared with lone tensor or vector GWs in the SGWB. Nonetheless, the results suggest that more precise measurements of the inter-pulsar correlation due to the SGWB might be able to close the window for this phenomenological model, similar with lone vector GWs forming the SGWB. This points out an interesting science prospect to pursue with future data.

Scalar modes superposing with tensor modes in the SGWB puts on display the need for more correlations data and the consideration of the CV (Figs. \ref{fig:HDphiconstraints}-\ref{fig:Tphiconstraints}). In these cases, neither the individual PTA or the joint analyses are able to give a constraint on the speed of the scalar modes and tensor modes in the T-$\phi$ model. Furthermore, the resulting total non-tensorial amplitude depends unreliably too much on whether the CV is considered. This is due to the scalar modes' dominant monopole contribution that manifests a monopolar mean signal and a relatively wide but flat CV with respect to the mean. This consequently makes the HD-$\phi$ and the T-$\phi$ correlation models overfit NANOGrav's measured correlations particularly when the CV is considered. The overfitting is reflected in the pointwise significance ($\overline{\chi}^2 \sim 0.1$) of both models and visually depicted by the quadrupolar belt in the second rows of Figs. \ref{fig:HDphiconstraints} and \ref{fig:Tphiconstraints}. The belt-like CV surrounding the tensor quadrupole that smoothly passes through the error bars of the NANOGrav correlation samples can be distinctly attributed to the monopolar power catered by the scalar modes (\cite{Qin:2020hfy, Bernardo:2022vlj, Bernardo:2022rif, Bernardo:2022xzl}).

The tighter CPTA data and by extension the joint analysis with the NANOGrav samples however give unreliable estimates to the phenomenological non-tensorial amplitude parameter. In particular, when the CPTA data is considered, the analysis returns an upper bound to the non-tensorial modes' fraction when the CV is not considered; otherwise, a lower bound is obtained when the CV is taken into account. This is contrary to the expectation that the CV should be considered and the scalar modes are subdominant to the tensor modes. In addition, no reliable information was obtained on the subluminal scalar/tensor modes' speed. These results merely point out the limitations of the present data and the importance of the CV for gravitational physics interpretations. This is especially true for forthcoming, more precise, data sets, as illustrated by the factoring in of the CPTA data in the above results.

Granted, there are already constraints to non-Einsteinian GWs from other observational probes such as ground-based GW detectors (\cite{Isi:2017fbj, LIGOScientific:2017ycc, LIGOScientific:2018czr}). However, these constraints only strictly apply to a frequency band that is orders of magnitude above from the cosmologically relevant nanohertz regime where gravity may be stranger. Posing independent nanohertz GW constraints on such GW modes is without doubt a desirable PTA science objective that complements the GW multiband picture.


This paper celebrates the pulsar timing array collaborations unwavering dedication to their craft and science mostly behind-the-scenes that led to the discovery of the stochastic gravitational wave background. 

The Hellings-Downs curve (\cite{Hellings:1983fr})---a theoretical predication based on general relativity, about 40 years ago---remains a perfectly acceptable model of the observed spatial correlations, even in the background of a variety of alternative models. This spectacular achievement once again goes to GR's incredible scientific resume (\cite{Taylor:1979zz, vanStraten:2001zk, Bertotti:2003rm, Ciufolini:2004rq, Valtonen:2008tx, Reyes:2010tr, ColemanMiller:2019tqn}).

In conclusion, the profound quadrupolar correlation observed in the timing residuals of millisecond pulsars serves as unequivocal evidence of the existence of a stochastic gravitational wave background. However, understanding the nature of this correlation in a gravitational framework demands a comprehensive exploration of alternative correlation scenarios and their compatibility with data. In this context, the Hellings-Downs curve emerges as a crucial reference point for investigation{, as supported by the Bayes factors in Tab. \ref{tab:summary}}. Our results shed light on the tantalizing possibility that the observed inter-pulsar correlations harbor deviations from the Hellings-Downs curve, underscoring an intimate interplay between gravity and pulsar timing arrays.

The work also brings to light several caveats that merit future consideration. Primary among these is the reliance on published inter-pulsar correlation data points, which were derived from different frequency bins. Our approach treated the uncertainty of the overall correlation measurement as indicative of alternative viable correlation models. While we anticipate that correlations resulting from relativistic degrees of freedom exhibit weak sensitivity to frequency, further investigation into the dependencies related to both frequency and pulsar distances is needed. A suggestion put forth for future nanohertz gravity investigation involves direct measurements of the timing-residual two-point function with actual pulsar distances, though that requires heavy computational power, or the power spectrum of the correlation. The latter avenue, facilitated by a likelihood in the power spectrum multipoles, may admit a more natural categorization into frequency bins, thereby mitigating the principal limitation associated with angular correlations.

A related limitation due to the reliance on published spatial correlations data in \cite{NANOGrav:2023gor} and \cite{Xu:2023wog} is that we cannot rigorously account for the fact that there are overlapping pulsars in both data sets. However, we have to wait until both data sets become public in order to assess the robustness of our joint analysis. Nonetheless, the spatial correlation is set apart compared with other noises in PTAs by the number of pulsar pairs that grow quadratically with the number of pulsars. This implies that the correction due to the overlapping pulsars in NG15 and CPTA is subdominant compared with the total number of pulsar pairs in the combined data set. We must content ourselves with this logic for the meantime, until it can be confirmed when the PTA data sets become publicly available.

The posterior shapes we found agree well with results by other independent groups (\cite{Wang:2023div, Bi:2023ewq, Chen:2024fir}). Statistical factors such as the size of the present data and the model complexity may chip in to explain unusual posterior shapes. The results are suggestive of future work looking into alternative, possibly non-Gaussian, likelihoods for constraining GW parameters with PTA data.

Another caveat involves the assumption that cosmic variance accurately represents the theoretical uncertainty. This comes from the fact that cosmic variance can be realized only when a sufficiently large number of pulsars is observed and when the stochastic field comprising the gravitational wave background adheres precisely to a Gaussian distribution. Both of these conditions require rigorous testing. Encouragingly, the current NANOGrav data already exhibits experimental uncertainty that is on par with the cosmic variance, providing some optimism that this trend may persist in future IPTA data. The Gaussianity hypothesis on the other hand is inherently linked to the source such that if it is cosmological, then the stochastic field would most likely be Gaussian. But again, this hypothesis also remains to be further tested.

\begin{acknowledgements}
      The authors thank Reinabelle Reyes for helpful comments on a preliminary draft. RCB is supported by an appointment to the JRG Program at the APCTP through the Science and Technology Promotion Fund and Lottery Fund of the Korean Government, and also by the Korean Local Governments in Gyeongsangbuk-do Province and Pohang City. RCB also acknowledges support from the NRF of Korea (Grant No. NRF-2022R1F1A1061590) funded by the Korean Government (MSIT). This work was supported in part by the National Science and Technology Council (NSTC) of Taiwan, Republic of China, under Grant No. MOST 112-2112-M-001-067. The authors also express special thanks to the Mainz Institute for Theoretical Physics (MITP) of the Cluster of Excellence PRISMA$^+$ (Project ID 39083149), for its hospitality and support that enabled academic exchanges that partially reflected to the present work.
\end{acknowledgements}

%
%

\begin{appendix}
\section*{Methods}

We use the public code \texttt{PTAfast} (\cite{2022ascl.soft11001B}) to generate the spatial correlation templates that we then compare with the observed inter-pulsar correlations of the NANOGrav (\cite{NANOGrav:2023gor}) and the CPTA (\cite{Xu:2023wog}) collaborations.

\texttt{PTAfast} is based on the power spectrum approach, also often referred to as harmonic analysis, that recasts the GW stochastic correlation, $\gamma_{ab}(\zeta)$, into a multipolar sum (\cite{Gair:2014rwa,Qin:2018yhy,Qin:2020hfy,Ng:2021waj,Bernardo:2022rif,Liang:2023ary, Nay:2023pwu}):
\begin{equation}
\label{eq:meancorr}
\gamma_{ab}\left(\zeta\right) = \sum_{l} \frac{2l+1}{4\pi} C_l P_l\left(\cos \zeta\right) \,,
\end{equation}
where the coefficients, $C_l$'s, are uniquely determined by the nature of the modes that enter the SGWB. In practice, the first few multipoles are referred to as the monopole ($l = 0$), dipole ($l = 1$), quadrupole ($l = 2$), octupole ($l = 3$) and quintupole ($l = 4$), with associated powers $l(l + 1) C_l$. Pictorially, depending on $l$, these multipoles contribute a $l/2$ turns in the range $\zeta \in [0, \pi]$; monopole is flat/no turn ($\gamma_{ab}(\zeta) = $ Constant), dipole has a half turn ($\gamma_{ab}(\zeta) \propto \cos \zeta$), quadrupole has full turn ($\gamma_{ab}(\zeta) \propto \cos (2\zeta)$), and so on. The cosmic variance of the stochastic signal admits a similarly simple power spectrum sum (\cite{Ng:1997ez, Allen:2022dzg, Bernardo:2022vlj}):
\begin{equation}
    \sigma^{\rm CV}_{ab}\left(\zeta\right) = \sum_l \frac{2l+1}{8\pi}C_l^2P_l\left(\cos\zeta\right) \,.
\end{equation}
This is a natural limit to precision for Gaussian fields such as the CMB.

Gravity shapes the $C_l$'s which are at its core a two point statistical correlation $C_l \propto \langle  r_a r_b \rangle \sim \langle  h_a h_b \rangle$ where $r_i$'s are the timing residual of pulsar $i$ induced by a passing gravitational wave $h_i$ (\cite{Bernardo:2022rif, Bernardo:2022xzl}). These are put beautifully by two equations,
\begin{equation}
\label{eq:powerspectrummultipoles}
    C_l^{\rm A} = {\cal F}_l^{\rm A}\left( fD,v \right) {\cal F}_l^{\rm A}\left( fD,v \right)^* / \sqrt{\pi}
\end{equation}
and
\begin{equation}
\label{eq:projectionfactors}
    \frac{{\cal F}_l^{\rm A}\left( y,v \right)}{N_l^{\rm A}} = \int_0^{\frac{2\pi y}{v_{\rm ph}}} dx \ v_{\rm ph} e^{ixv_{\rm ph}} \frac{d^q}{dx^q} \left( \frac{j_l(x) + r v_{\rm ph}^{-2} j_l''(x)}{x^p} \right) \,,
\end{equation}
where $j_l(x)$'s are the spherical Bessel functions, $v_{\rm ph}$ is the phase velocity, $v$ is the GW speed, $f$ is the GW frequency, $D$ are the pulsar distances. The coefficients $N_l^{\rm A}$ and indices $p, q, r$ are given in Table \ref{tab:powerspectrumcoefs}, beautifully summarizing a decades' theoretical work that went to the analysis of non-Einsteinian GW correlations.
\begin{table}[h!]
\centering
\caption{The SGWB angular power spectrum coefficients and indices (equations \ref{eq:powerspectrummultipoles} and \ref{eq:projectionfactors}) for arbitrary GW modes.}
\label{tab:powerspectrumcoefs}
\renewcommand{\arraystretch}{1.25}
\begin{tabular}{|r|r|r|r|r|}
\hline
\ GW mode \ & $N_l^{\rm A}/\left(2 \pi i^l\right)$ \ & \phantom{g} $p$ \ & \phantom{g} $q$ \ & \phantom{g} $r$ \ \\
\hline 
Tensor \ & \phantom{g} $\sqrt{(l+2)!/(l-2)!}/\sqrt{2}$ \ & 2 \ & 0 \ & 0 \ \\ \hline
Vector \ & $\sqrt{2} \sqrt{l(l + 1)}$ \ & 1 \ & 1 \ & 0 \ \\ \hline
Scalar \ & $1$ \ & 0 \ & 0 \ & 1 \ \\ \hline
\end{tabular}
\end{table}

\texttt{PTAfast} assumes $v = v_{\rm g} = 1/v_{\rm ph}$ which is valid whenever the massive dispersion relation, $\omega^2 = k^2 + \mu^2$, holds in weak fields such as the tensor GWs in massive gravity and the scalar modes in Horndeski theory. Other theories can have a different relation, $v_{\rm g} = v_{\rm g}\left(v_{\rm ph}\right)$, depending on their underlying dispersion relation (\cite{Liang:2023ary}).

For scalar GWs, the integration simplifies to zero for $l \geq 2$ in the limit $fD\rightarrow \infty$ (\cite{Qin:2020hfy, Bernardo:2022vlj}). At distances characterizing the PTA millisecond pulsars, $fD \sim {\cal O}\left(10^{2-3}\right)$ at $f \sim {\cal O}(10^{1-2})$ nanohertz, propagating scalars are thus distinguished by their monopole and dipole contributions. Vector GWs on the other hand are characterized by a dominant quadrupole and higher multipoles with the addition of a dipole. Tensor GWs are characterized by a quadrupole followed by suppressed higher moments. The Hellings-Downs correlation is a particular case of tensor modes in the infinite distance and luminal speed limit (\cite{Gair:2014rwa, Ng:2021waj}),
\begin{equation}
    C_l^{\rm HD} \sim \frac{8 \pi^{3/2}}{(l-1)l(l+1)(l+2)} \,\,\,\,, \, l \geq 2 \,.
\end{equation}
The correlation function $\gamma_{ab}(\zeta)$ is normalized to the overlap reduction function GW templates, $\Gamma_{ab}(\zeta)$, traditionally normalized with respect to the HD correlation as $\Gamma_{ab}(\zeta) = 0.5 \gamma_{ab}(\zeta)/\gamma_{ab}(0)^{\rm HD}$, so that $\Gamma_{ab}^{\rm HD}(0) = 0.5$. These are straightforwardly calculated and returned by \texttt{PTAfast}.

To sample the parameter space of the correlation models, we utilize the nested sampler polychord (\cite{Handley:2015fda, 2015MNRAS.453.4384H}) anchored in the cosmology community code \texttt{Cobaya} (\cite{Torrado:2020dgo}) and analyze the statistical results using \texttt{GetDist} (\cite{Lewis:2019xzd}). We consider a Gaussian likelihood for the inter-pulsar correlations,
\begin{equation}
    {\log \mathcal{L}} \propto {-\frac{1}{2}} \sum_{{\text{pulsar pairs}}} \left( \frac{ \Gamma_{ab} ^{\rm data} - \Gamma_{ab}^{\rm model} }{\Delta \Gamma_{ab}^{\rm data}} \right)^{{2}} \,,
\end{equation}
where the correlations data are given as $\Gamma_{ab}^{\rm data} \pm \Delta \Gamma_{ab}^{\rm data}$. The cosmic variance are considered in the analysis as $\Delta \Gamma_{ab}^{\rm eff} = \sqrt{\left( \Delta \Gamma_{ab}^{\rm data} \right)^2 + \sigma_{ab}^{\rm CV}}$. The samplings are performed with flat priors for the correlation model parameters: the speed $v \in [0, 1]$ and the non-tensor fraction $r^2 \in [0, 1]$.

For the tensor-non-tensor phenomenological correlation models (HD-V, HD-$\phi$, and T-$\phi$), we consider an effective overlap reduction function and cosmic variance,
\begin{equation}
\label{eq:twopoint_tnt}
    \Gamma_{ab}(\zeta) = \Gamma_{ab}^{\rm Tensor}(\zeta) + r^2 \Gamma_{ab}^{\rm Nontensor}(\zeta) 
\end{equation}
and
\begin{equation}
\label{eq:cosmicvariance_tnt}
    \sigma_{ab}^{\rm CV}(\zeta) = \sigma_{ab}^{\rm CV,Tensor}(\zeta) + r^4 \sigma_{ab}^{\rm CV,Nontensor}(\zeta) \,.
\end{equation}
Recognizing that $\Gamma_{ab} \sim \langle h_a h_b \rangle$ and $\sigma^{\rm CV}_{ab} \sim \langle h_a h_a h_b h_b \rangle$ are two- and four-point functions of the gravitational wave amplitude (\cite{Bernardo:2022xzl}), the phenomenological parameter $r^2$ could be associated with the fraction of non-tensorial modes in the superposition.

A physical way to realize this is by recognizing that the pulsar timing residual due to GW (at time $t$ for a pulsar in the direction $\hat{e}$ along the light of sight) can be written as a harmonic series,
\begin{equation}
    R\left(t, \hat{e}\right)=\sum_{lm} a_{lm}(t) Y_{lm}\left(\hat{e}\right) \,,
\end{equation}
where the coefficients,
\begin{equation}
    a_{lm}(t)=2\pi \sum_{A}\int df d\hat{k} \ F\left(t, f, D, v, \hat{k} \right) \left( \hat{e}^i \otimes \hat{e}^j \right) \varepsilon_{ij}^A h_{A}\left(f, \hat{k}\right) \,,
\end{equation}
embody the stochastic superposition of GWs of different frequencies $f$, directions $\hat{k}$, and polarizations $A$ (see \cite{Bernardo:2022rif} for the full expression). Central to the point we wish to make is that the sum over polarizations can be broken down to
\begin{equation}
    a_{lm}(t) = ({\rm Other \ factors}) \times \left( \sum_{T} \varepsilon_{ij}^T h_{T}\left(f, \hat{k}\right) + r \sum_{\cancel{T}} \varepsilon_{ij}^{\cancel{T}} h_{\cancel{T}}\left(f, \hat{k}\right) \right) \,,
\end{equation}
where $T$ and $\cancel{T}$ stand for the tensor and non-tensorial polarization components of the GW superposition, respectively, and $r$ is a constant manifested through non-tensorial GWs, e.g., $r = 0$ means that there are only tensorial GWs. Thus, provided that the SGWB is Gaussian and isotropic, i.e.,
\begin{equation}
    \langle h_{A}\left(f, \hat{k}\right) h_{A'}\left(f', \hat{k}'\right) \rangle = P_A\left(f\right) \delta_{AA'} \delta(f-f')\delta\left(\hat{k}-\hat{k}'\right) \,,
\end{equation}
we see that in computation the two point function of the timing residuals of a pair of pulsars, $\langle R\left( t, \hat{e}_a \right) R\left( t, \hat{e}_b \right) \rangle$, the independence of power between tensorial and non-tensorial GW polarizations, manifested by the Kronecker delta, $\delta_{T\cancel{T}}$, naturally leads to the spatial correlation \eqref{eq:twopoint_tnt} and \eqref{eq:cosmicvariance_tnt} with the constant $r$ appearing as the phenomenological quantity that we associated with non-tensorial modes.

Graviton masses are computed by inverting the massive dispersion relation and using the GW group speed $v = d\omega/dk$, leading to the approximate expression $m_{\rm g} \simeq \left( 1.3105 \times 10^{-22} \ {\rm eV} \right) \times f\left[{\rm yr}^{-1}\right] \times \sqrt{1 - v^2}$ for the nanohertz GW band.

Python notebooks and codes that were used to obtain the results of this work can be downloaded in the \href{https://github.com/reggiebernardo/PTAfast}{\texttt{PTAfast} repository}.
    
\end{appendix}

\end{document}